\documentclass[12pt]{iopart}

\usepackage{amsbsy}
\usepackage[english]{babel}
\usepackage{graphicx}
\usepackage{units}

\begin{document}

\title[]{Kondo screening in high-spin side-coupled two-impurity clusters}

\author{R \v{Z}itko}
\address{Jo\v{z}ef Stefan Institute, Jamova 39, SI-1000 Ljubljana, Slovenia}
\ead{rok.zitko@ijs.si}

\date{\today}

\begin{abstract}
We study the system of two magnetic impurities described by a two-impurity
Kondo model where only the first impurity couples directly to the conduction
band, while the second impurity interacts with the first through Heisenberg
exchange coupling in a ``side-coupled'' configuration. We consider various
choices of the impurity spins ($S_1<S_2$, $S_1=S_2$, and $S_1>S_2$) and we
contrast the regimes where the inter-impurity exchange coupling $J$ is
either lower or higher than the Kondo temperature $T_K^{(0)}$ of the first
impurity in the absence of the second. This model is a high-spin
generalization of the two-impurity model for side-coupled double quantum
dots which corresponds to the simplest $S_1=S_2=1/2$ case, where the moments
are Kondo screened successively in two stages for $J<T_K^{(0)}$ (the
``two-stage Kondo effect''). We show that the two-stage Kondo screening
occurs generically for $S_2 \geq S_1$. For $S_1 \geq 1$, the second Kondo
temperature $T_K^{(2)}$ is not exponentially reduced, as for $S_1 = 1/2$,
but is approximately a power-law function of the coupling $J$. Furthermore,
for $S_1 \geq 1$ all three scales ($T_K^{(0)}$, $J$, $T_K^{(2)}$) explicitly
appear in the temperature-dependence of the thermodynamic properties. For
$S_1 > S_2$, there is no second stage of screening for $J<T_K^{(0)}$,
however in the opposite limit $J>T_K^{(0)}$ the Kondo screening of the
effective spin $S_1-S_2$ is found.
\end{abstract}

\pacs{72.10.Fk, 72.15.Qm}

\maketitle

\newcommand{\vc}[1]{{\boldsymbol{#1}}}
\newcommand{\ket}[1]{|#1\rangle}
\newcommand{\bra}[1]{\langle #1|}
\newcommand{\braket}[1]{\langle #1 \rangle}
\renewcommand{\Im}{\mathrm{Im}}
\renewcommand{\Re}{\mathrm{Re}}
\newcommand{\dr}{\mathrm{d}}
\newcommand{\correl}[1]{\langle\langle #1 \rangle\rangle_\omega}
\newcommand{\TKO}{T_K^{(0)}}
\newcommand{\TKt}{T_K^{(2)}}

\newcommand{\eqref}[1]{\eref{#1}}

\bibliographystyle{unsrt}

\section{Introduction}

When magnetic impurities, such as substitutional defects in the bulk or
adatoms on the surface, couple with the conduction band electrons through an
antiferromagnetic exchange interaction, their spin is screened in the Kondo
effect and the level degeneracy is effectively lifted \cite{hewson,
anderson1970, nozieres1974, nozieres1980, andrei1983, tsvelick1983,
gunnarsson1983a}. When the separation between two such impurities is small,
the impurities interact through the RKKY interaction \cite{ruderman1954},
which may lead to critical behaviour in some parameter regimes
\cite{jones1989prb, affleck1992}. A simplified description of such systems
is the two-impurity Kondo model \cite{jayaprakash1981, jones1987, jones1988,
jones1989, sakai1992, silva1996}: the two magnetic atoms are represented by
quantum spin operators which are coupled by some exchange interaction $J$,
and each furthermore interacts with the conduction band electrons though an
effective Kondo exchange coupling. With few exceptions \cite{lehur1997,
vojta2002}, most studies of such models focus on spin-$1/2$ impurities,
while real impurities may actually have higher spin \cite{nozieres1980}. The
same may also be the case in artificial atoms, i.e. quantum dots
\cite{sasaki2000}, and in molecules with embedded magnetic ion
\cite{tsukahara2009}.  Due to competing interactions and the vastness of the
parameter space, a great variety of different types of magnetic behavior are
expected. In this work we discuss a sub-class of high-spin two-impurity
models in which only one of the spins ($S_1$) couples to the
conduction-band, while the second spin ($S_2$) is ``side-coupled'' to the
first one. Only the $S_1=S_2=1/2$ limit of this family has been studied so
far \cite{vojta2002}, and some results are known for the case of $S_1=1/2$
and arbitrary $S_2$ in the related Anderson-Kondo model
\cite{peters2006fluc}. It is shown that the two-stage Kondo screening
\cite{vojta2002, cornaglia2005tsk, sidecoupled} found in the $S_1=S_2=1/2$
model is a generic feature of all $S_2 \geq S_1$ models, although for $S_1
\geq 1$ some qualitative differences arise.

\section{Model and method}

We consider the two-impurity Kondo model
\begin{equation}
\label{H}
H = \sum_{k\sigma} \epsilon_k c^\dag_{k\sigma} c_{k\sigma} 
+ J_K \vc{s} \cdot \vc{S}_1 + J \vc{S}_1 \cdot \vc{S}_2 
\end{equation}
Operators $c^\dag_{k\sigma}$ create conduction band electrons with momentum
$k$, spin $\sigma \in \{\uparrow,\downarrow\}$, and energy $\epsilon_k$,
while $\vc{s}=\{ s_x, s_y, s_z \}$ is the spin density of the
conduction-band electrons at the position of the first impurity. Operators
$\vc{S}_1 = \{ S_{1,x}, S_{1,y}, S_{1,z} \}$ and $\vc{S}_2 = \{ S_{2,x},
S_{2,y}, S_{2,z} \}$ are the quantum-mechanical impurity spin operators.
Furthermore, $J_K$ is the effective Kondo exchange coupling constant, and
$J$ is the inter-impurity Heisenberg coupling constant. The models
considered are the simplest generalization of the spin-$1/2$ two-impurity
models for the side-coupled impurity configuration exhibiting the two-stage
Kondo effect \cite{vojta2002, cornaglia2005tsk, sidecoupled, sidecoupled2},
which can also be found in other multi-impurity problems \cite{flnfl3,
trikotnik}. Two-stage Kondo screening occurs when the exchange coupling
between the impurities is weaker than the energy scale of the Kondo
screening of the directly coupled impurity. In such situation, the moment on
the first impurity is Kondo screened at the Kondo temperature which
corresponds to a decoupled impurity, $T_K^{(0)}$, while the moment on the
second is Kondo screened at some exponentially reduced temperature
$T_K^{(2)}$. A simple interpretation is that the second Kondo effect occurs
due to exchange coupling between the side-coupled impurity and the Fermi
liquid of heavy quasiparticles resulting from the first stage of the Kondo
screening \cite{cornaglia2005tsk}.

We solve the Hamiltonian using the numerical renormalization group (NRG)
method \cite{wilson1975, krishna1980a, bulla2008}. In this approach, the
continuum of the conduction-band electron states is discretized logarithmically
with increasingly narrow intervals in the vicinity of the Fermi level, the
problem is transformed (tridiagonalized) to the form of a tight-binding
Hamiltonian with exponentially decreasing hopping constant which is then
diagonalized iteratively. A very effective technique to examine the magnetic
behavior of an impurity model consists in studying the thermodynamic
properties of the model as a function of the temperature \cite{wilson1975,
krishna1980a, krishna1980b}; the impurity contribution to the entropy,
$S_\mathrm{imp}$, then provides information on the degeneracy of the
effective spin multiplets, and the impurity contribution to the magnetic
susceptibility, $\chi_\mathrm{imp}$, defines the effective magnetic moment.

The results reported in this work have been calculated with the
discretization parameter $\Lambda=2$ using improved discretization
schemes \cite{campo2005, resolution}, without the $z$-averaging, and
with the NRG truncation cutoff set at $7\omega_N$, where $\omega_N$ is
the characteristic energy scale at the $N$-th step of the iteration.

\section{Properties of two antiferromagnetically coupled isotropic spins}
\label{spinprop}

We first briefly review some properties of a decoupled pair of magnetic
impurities described as pure spins with spin quantum numbers $S_1$ and
$S_2$; without loss of generality, in this section we use the convention
that $S_2 \geq S_1$. The results will be relevant in the discussion of the
$J \to \infty$ limit. 

The impurities couple antiferromagnetically into an effective $S=S_2-S_1$
spin object. The inner-product of spin operators is given as 
\begin{equation}
\braket{\vc{S}_1 \cdot \vc{S}_2}=-S_1(S_2+1).
\end{equation}
The ground state multiplet can be expressed using the Clebsch-Gordan
coefficients as
\begin{equation}
\ket{S,M} = \sum_{m_1,m_2} \braket{S_1,m_1,S_2,m_2|S,M} \ \ \ket{S_1,m_1} 
\otimes \ket{S_2, m_2}.
\end{equation}
The projections of the spin-$S$ object on the constituent spin operators can
be easily computed as the expectation values of the $S_{z,i}$ operators in
the maximum weight states $\ket{S,S}$:
\begin{eqnarray}
p_1 &= \bra{S,S}S_{z,1}\ket{S,S} \\
&= \sum_{m_1}
\left| \braket{S_1,m_1,S_2,S_2-S_1-m_1|S_2-S_1,S_2-S_1} \right|^2 m_1 \\
&= \frac{S_1 (S_2-S_1)}{S_1-S_2-1},
\end{eqnarray}
and
\begin{eqnarray}
p_2 &= \bra{S,S}S_{z,2}\ket{S,S} \\
&= \sum_{m_2}
\left| \braket{S_1,m_1,S_2,S_2-S_1-m_1|S_2-S_1,S_2-S_1} \right|^2 (S_2-S_1-m_1) \\
&= \frac{(S_1-S_2)(1+S_2)}{S_1-S_2-1}.
\end{eqnarray}
If the Hamiltonian describing the coupling of the impurities with the host
conduction band is of the form
\begin{equation}
H_C=J_1 \vc{S}_1 \cdot \vc{s} + J_2 \vc{S}_2 \cdot \vc{s},
\end{equation}
then the coupling of the effective spin takes the following form
\begin{equation}
H_C^{\mathrm{eff}}=J_\mathrm{eff} \vc{S} \cdot \vc{s},
\end{equation}
with
\begin{equation}
J_\mathrm{eff} = r_1 J_1 + r_2 J_2,
\end{equation}
where $r_i = p_i/(S_2-S_1)$. The ratios $r_i$ are thus the
multiplicative factors which determine the effective Kondo exchange
coupling of the composite object; they are tabulated in
Table~\ref{table1}. Note that the signs of $r_1$ is always negative,
while the sign of $r_2$ is always positive. This implies that in the
side-coupled configuration discussed in this work, the Kondo screening
of the effective spin in the $J \to \infty$ limit occurs only if
the impurity which couples to the conduction band is the one with
larger spin; in this case the impurity ground-state multiplet will
have spin $|S_2-S_1|-1/2$. In the opposite case, the exchange coupling
to the conduction band is ferromagnetic and the ground-state multiple
will have spin $|S_2-S_1|$.

\begin{table}[htb]
\caption{\label{table1}
Multiplicative factors which determine the effective Kondo exchange coupling
of the composite object made of the two spins locked into a $S=S_2-S_1$
antiferromagnetically aligned state.
}
\begin{indented}
\item[]\begin{tabular}{@{}llll@{}}
\br
$S_1$ & $S_2$ & $r_1$ & $r_2$ \\
\mr
$1/2$ & $1$   & $-1/3$ & $4/3$ \\
$1/2$ & $3/2$ & $-1/4$ & $5/4$ \\
$1/2$ & $2$   & $-1/5$ & $6/5$ \\
$1$   & $3/2$ & $-2/3$ & $5/3$ \\
$1$   & $2$   & $-1/2$ & $3/2$ \\
$3/2$ & $2$   & $-1$   & $2$   \\
\br
\end{tabular}
\end{indented}
\end{table}

\section{Results}

We fix $\rho J_K = 0.1$ throughout this work. The Kondo temperature of a
decoupled impurity is thus $T_K^{(0)} = 1.16 \times 10^{-5}W$ (Wilson's
definition) and it is the same for any value of spin $S_1$ \cite{rajan1982,
vzporedne, aniso}.

\begin{figure}[htbp]
\centering
\includegraphics[width=14cm,clip]{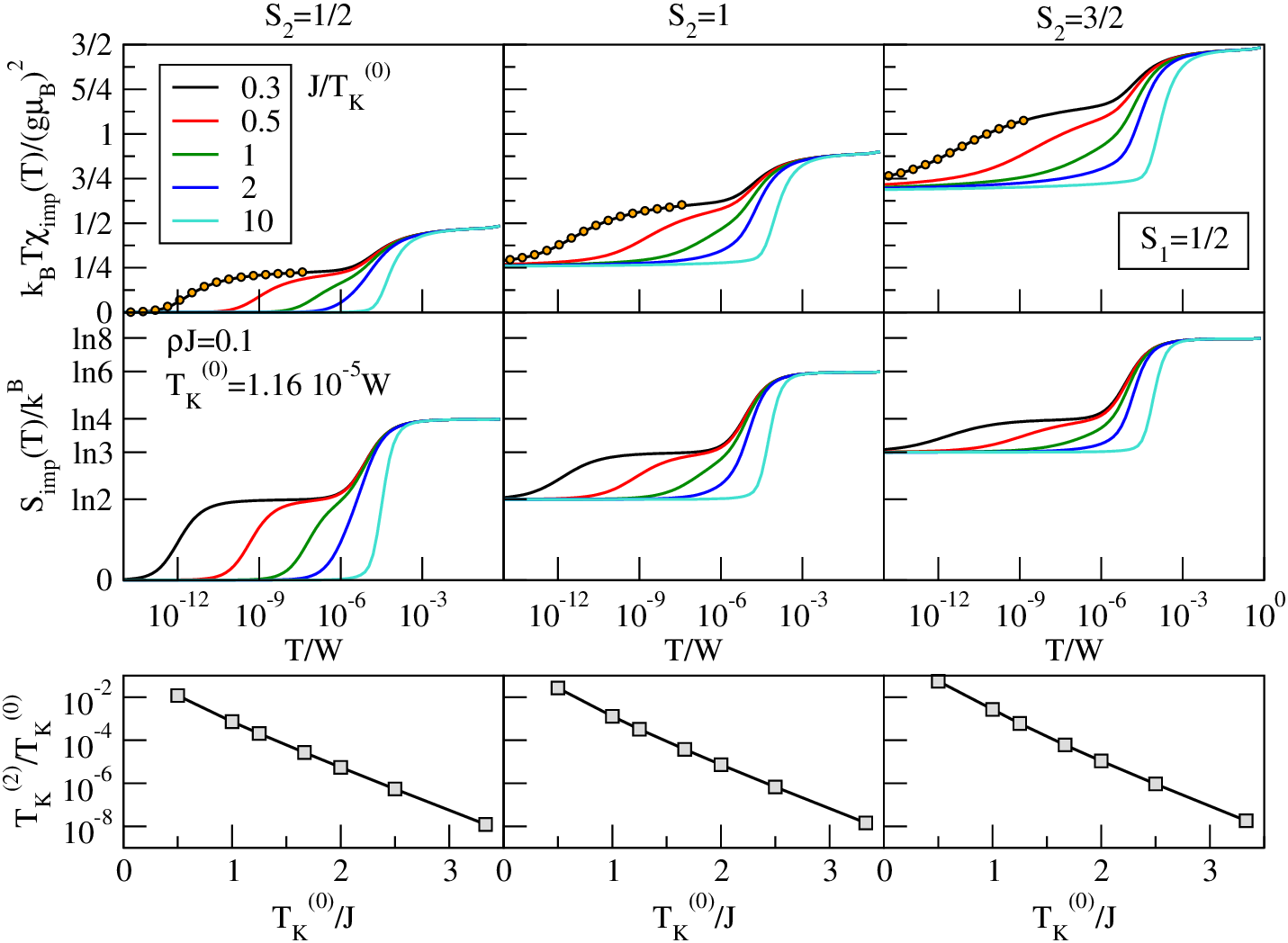}
\caption{Thermodynamic properties of the isotropic two-impurity clusters
with $S_1=1/2$. We plot the impurity contribution to the magnetic
susceptibility, $\chi_\mathrm{imp}$, and the impurity contribution to the
entropy, $S_\mathrm{imp}$. The symbols indicate fits to the universal
spin-$S_2$ Kondo magnetic susceptibility curves, which were used to extract
the $J$-dependence of the second Kondo temperature, displayed in the bottom
line of panels.}
\label{fig1}
\end{figure}

The thermodynamic properties of the system for the case when the first spin
is $S_1=1/2$ are shown in \Fref{fig1}. For $S_2=1/2$ we recover exactly the
prototypical two-stage Kondo screening where the second Kondo temperature is
given by \cite{vojta2002, cornaglia2005tsk}
\begin{equation}
T_K^{(2)} = c_1 T_K^{(1)} e^{-c_2 \frac{T_K^{(1)}}{J}},
\end{equation}
where $c_1$ and $c_2$ are some numeric constants; the scaling of $T_K^{(2)}$
with $T_K^{(0)}/J$ is shown in the subfigure in the bottom panel.

For $S_2 \geq 1$, we observe very similar behaviour: for $J < T_K^{(0)}$,
after the initial screening of the first impurity, the second impurity
undergoes spin-$S_2$ Kondo screening which reduces its spin by one
half-unit, as if it were coupled directly to the conduction band. The only
effect of the first impurity is thus to induce much lower effective
bandwidth $D_{\mathrm{eff}} \propto T_K^{(0)}$ and increased density of
states $1/\rho_{\mathrm{eff}} \propto T_K^{(0)}$. This picture is confirmed
by the scaling of $T_K^{(2)}$ with $T_K^{(0)}/J$ which is very similar for
all three values of $S_2$ in \Fref{fig1}.

\begin{figure}[htbp]
\centering
\includegraphics[width=14cm,clip]{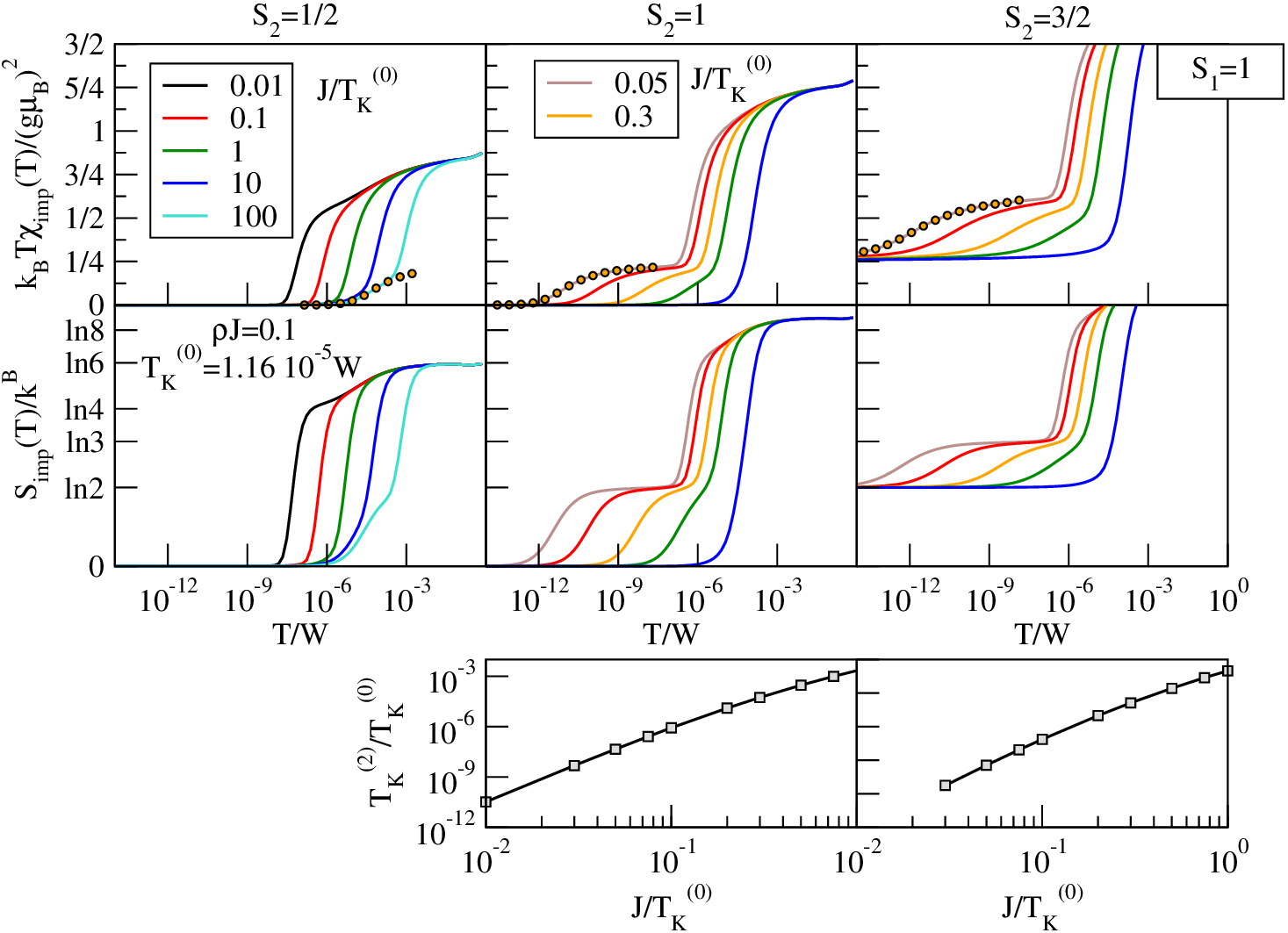}
\caption{Thermodynamic properties of the two-impurity clusters with $S_1=1$.
For $S_2=1$ and $S_2=3/2$, the scaling of the second Kondo temperature with
$J$ is shown in the bottom panels; it should be noted that the horizontal
axis here corresponds to $J/T_K^{(0)}$ and that the scale is logarithmic,
while in \fref{fig1} the axis corresponded to the inverse $T_K^{(0)}/J$,
while the scale was linear. The symbols for $S_2=1/2$ and $S_2=1$
corresponds to fits using the universal spin-$1/2$ Kondo magnetic
susceptibility, while the symbols for $S_2=3/2$ correspond to a fit using
the universal spin-$1$ Kondo magnetic susceptibility.
}
\label{fig2}
\end{figure}

We now consider the case when the first spin is $S_1=1$. We remind the
reader that when a single impurity with spin $S \geq 1$ couples to the
conduction band, one half-unit of the spin is screened in a spin-$S$ Kondo
effect, giving rise to a spin $S-1/2$ composite object which couples with
the conduction band electrons with a ferromagnetic effective exchange
coupling, thus it remains unscreened in the ground state \cite{mattis1967,
cragg1979b, cragg1980, koller2005, mehta2005, vzporedne, borda2009}. The
situation becomes more involved in the presence of an additional impurity,
see \Fref{fig2}.

For $S_1=1$ and $S_2=1/2$, in the limit $J \gg T_K^{(0)}$ the two spins
couple into a spin-$1/2$ object which couples to the conduction band with an
antiferromagnetic effective Kondo exchange coupling $J_\mathrm{eff}=r_1 J$
with $r_1=4/3$ (see \Tref{table1}) and undergoes the usual spin-$1/2$ Kondo
effect, which results in fully compensated impurity spins and a
non-degenerate singlet ground state. The expression for $J_\mathrm{eff}$
holds strictly only when $J$ is much larger than any other scale in the
problem (in particular $J_K$ and the band-width $W$); for $J \approx 60 J_K$
we indeed find that the Kondo temperature is $T_K = 1.7 \times 10^{-4} W$
which agrees with the expected scale of $T_K \approx W \sqrt{\rho J (4/3)}
\exp\{-1/[\rho J (4/3)]\} \approx 2 \times 10^{-4} W$. For lower $J$, the
effective band-width is of order $J$ rather than $W$, thus the Kondo
temperature is reduced accordingly. An example of such behaviour is shown in
\Fref{fig2} for $J/T_K^{(0)}=100$. The initial free-spin $\ln 6$ entropy
entropy is reduced to $\ln 2$ at $T \sim J$ upon formation of the effective
composite spin. This is followed by the conventional spin-$1/2$ Kondo
screening (see the fit of the magnetic susceptibility with the universal
Kondo curves).

In the opposite limit $J \ll T_K^{(0)}$ we observe the initial spin-$1$
Kondo screening of the first spin: the magnetic susceptibility goes toward
$1/2$ and the entropy toward $2\ln2$. The screening is, however, abruptly
interrupted at $T \sim J$. This can be interpreted as the formation of a
spin-singlet object composed from the residual spin-$1/2$ resulting from the
Kondo screening and the side-coupled spin-$1/2$ impurity. The end result is
the same in both large-$J$ and small-$J$ limits; the cross-over between the
two is smooth as a function of $J$.

For $S_1=1$ and $S_2=1$, the behaviour for $J \gg T_K^{(0)}$ is particularly
simple, since the two spins bind at the temperature $T \sim J$ into a
singlet and they no longer play any role. For $J \ll T_K^{(0)}$ the Kondo
screening of the $S_1=1$ spin into a residual spin-$1/2$ is interrupted at
the temperature $T \sim J$. The residual spin-$1/2$ then binds with the
side-coupled $S_2=1$ into a new spin-$1/2$ composite object. Unlike the
residual spin-$1/2$ resulting from the incomplete screening of a spin-$1$
Kondo impurity, which remains uncompensated since it couples to the
conduction band ferromagnetically, the spin-$1/2$ composite object that
emerges in this case couples with the conduction band antiferromagnetically,
thus at some lower temperature which we again denote $T^{(2)}_K$ it is
compensated in a spin-$1/2$ Kondo effect. This thus constitutes a
non-trivial generalization of the two-stage Kondo screening phenomenology
encountered in the $S_1=1/2$ cases. The differences, however, are notable:
1) there are not two, but three energy scales: $T_K^{(0)}$, where the
spin-$1$ Kondo screening takes place, $J$, where this screening is abruptly
interrupted, and $T^{(2)}_K$, where the second Kondo screening occurs; 2)
the scaling of the second Kondo temperature $T^{(2)}_K$ is not exponential
with $1/J$. In the conventional two-stage Kondo effect with $S_1=1/2$, the
only role of the coupling $J$ is to set the lower Kondo temperature; no
feature is observed there in the thermodynamic properties of the system at
$T \sim J$. Here, the coupling $J$ is essential to produce a composite spin
object which then couples antiferromagnetically with the rest of the system,
thus this scale is directly observable as a sharp change in the effective
impurity degrees of freedom at $T \sim J$. The second Kondo temperature is
defined by a power-law with the exponent near 3, with some corrections (see
the lower panels in \Fref{fig2}).

For $S_1=1$ and $S_2=3/2$ the results for $J \gg T_K^{(0)}$ are trivial: at
the temperature $T \sim J$, the spins lock into a spin-$1/2$ object which
couples ferromagnetically with the conduction band, thus the composite spin
remains unscreened. This is in accord with the expected behavior in this
limit (see \sref{spinprop}). For $J \ll T_K^{(0}$ the results are, however,
very intriguing: the Kondo screening of the $S_1=1$ spin is interrupted at
the temperature $T \sim J$. At this point, the residual spin-$1/2$ couples
antiferromagnetically with the $S_2=3/2$ spin into a spin-$1$ composite
object. This composite object, interestingly, couples antiferromagnetically
with the conduction-band electrons, which leads to Kondo screening of one
half-unit of spin at some lower temperature which we denote, yet again, as
$T^{(2)}_K$. The final residual spin-$1/2$ is not compensated, since it
couples ferromagnetically with the conduction band. Thus we again observe a
two-stage Kondo effect of the same universality class as in the $S_1=S_2=1$
case. This result may, in fact, be generalized: for any $S_2 \geq S_1$, the
impurity spins will be compensated in two screening stages (the compensation
is only partial for $S_2 \neq S_1$).

\begin{figure}[htbp]
\centering
\includegraphics[width=14cm,clip]{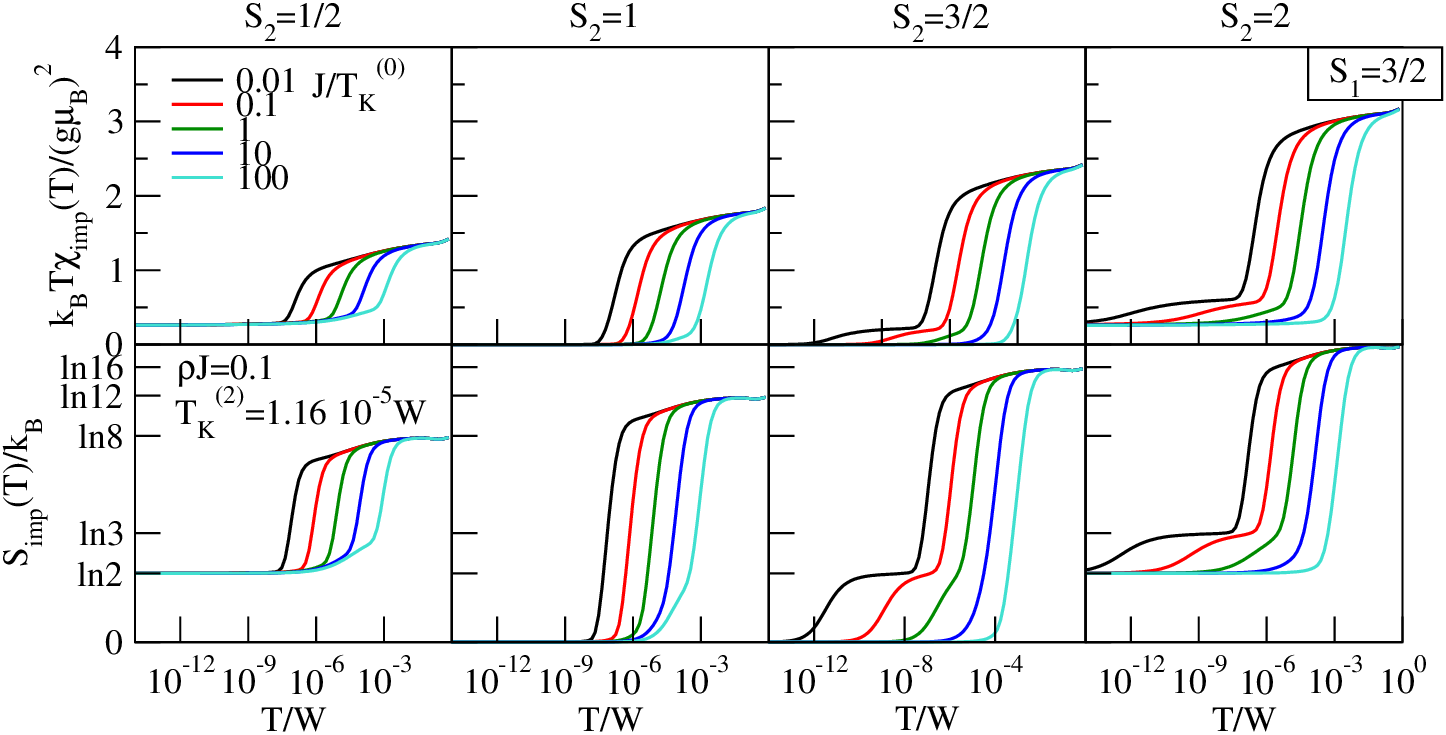}
\caption{Thermodynamic properties of the two-impurity clusters with
$S_1=3/2$. For $S_2=1/2$, the Kondo screening in the large-$J$ limit is of
the spin-$1$ type, while for $S_2=1$ the Kondo screening in the same limit
is of the spin-$1/2$ type. For $S_2=3/2$, the second stage Kondo screening
in the small-$J$ limit is of the spin-$1/2$ type, while for $S_2=2$ it is of
the spin-$1$ type.}
\label{fig3}
\end{figure}

To further substantiate the claim that the results are generic, we show in
\Fref{fig3} the results for the $S_1=3/2$ case. For $S_2=1/2$ and $S_2=1$,
i.e. for $S_2 < S_1$ we again find the emergence of the Kondo screening of
the rigidly antiferromagnetically bound $S_2-S_1$ spin in the large-$J$
limit, and the formation of a $S=(S_2-1/2)-S_1$ bound state at $T \sim J$ in
the small-$J$ limit. Furthermore, in the case of $S_2 \geq S_1$, the
two-stage Kondo screening is again observed in the small-$J$ limit, again
with power-law dependence of $T_K^{(2)}$ on $J$.

\section{Discussion and conclusion}

We have shown that when a second impurity is side-coupled to a Kondo
impurity with sufficiently small Heisenberg coupling $J$, the spin will be
screened in two stages for all systems, where the spin of the side-coupled
impurity $S_2$ is equal or greater than the spin of the directly coupled one
$S_1$. When $S_1=1/2$, the second Kondo temperature is exponentially
reduced, while for $S_1 \geq 1$, it is a power-law function of the coupling
$J$. The difference stems from the fact that for $S_1=1/2$, the second stage
of the Kondo screening occurs with a {\sl local} spin $S_2$ which interacts
with a Fermi liquid of heavy electrons resulting from the first screening
stage, while for $S_1 \geq 1$ the first screening stage leaves behind a
residual uncompensated spin $S_1-1/2$, which is an {\sl extended} object.
This residual spin then rigidly binds with the spin of the side-coupled
impurity at the temperature scale of $T \approx J$ to produce a new {\sl
extended} spin object which then undergoes Kondo screening. Similar
behaviour is found in the anisotropic single-impurity Kondo model, where an
easy-plane anisotropy leads to a formation of an extended effective
spin-$1/2$ degree of freedom which is Kondo screened \cite{aniso}; in this
problem, the second Kondo temperature is a power-law function of the
longitudinal magnetic anisotropy constant $D$. No theory has been devised
yet to map this class of problems with effective extended spin degrees of
freedom onto the conventional Kondo model with a localized spin operator,
thus there is presently no analytical account of these power-law
dependences. Nevertheless, it is clear that a power-law dependence of the
second Kondo temperature, i.e., $T_K^{(2)} \propto \exp(-1/\rho_\mathrm{eff}
J_\mathrm{eff}) \propto J^\alpha$, implies an inverse logarithmic dependence
of the effective impurity parameters, i.e., $\rho_\mathrm{eff}
J_\mathrm{eff} \propto -1/\ln J$. This form is suggestive of the energy
dependence of the renormalized ferromagnetic exchange coupling of the
residual spin in the underscreened Kondo model, ${\tilde
J}(\omega)=1/\ln(\omega/T_0)$, where $T_0$ is some low energy scale
\cite{koller2005}. This indicates that the ferromagnetic residual coupling
might play a decisive role in determining the total effective
antiferromagnetic exchange coupling of the composite spin object. In this
scenario, the bare parameter $J$ leads to the emergence of the composite
spin object by antiferromagnetic binding of the residual spin with the
side-coupled impurity spin, which occurs on the energy scale of $\omega=J$,
while the coupling of this object with the surrounding electron liquid is
controlled solely by ${\tilde J}(\omega=J)=1/\ln(J/T_0)$. This may be
explained by the fact that the side-coupled impurity interacts with the
electron liquid only indirectly through the first impurity and that the sign
of the relevant multiplicative factor $r_1$ is always negative (see
Table~\ref{table1}), thus the sign of the effective exchange interaction is
flipped.

In conclusion it may also be remarked that a common feature of all models
considered in this work is that the ground state in no way depends on the
$J/T_K^{(0)}$ ratio; for any non-zero Heisenberg coupling between the
impurities we always end up in the same fixed point, only the temperature
dependence of the spin-compensation differs greatly. This no longer holds
for problems with additional magnetic anisotropy terms (i.e., two-impurity
extensions of models studied in Refs.~\cite{aniso, aniso2}), where level
crossings may also occur as a function of $J$. This behavior will be
addressed in future works.

\ack
The author acknowledges the support of the Slovenian Research Agency
(ARRS) under Grant No. Z1-2058.

\bibliography{paper}

\end{document}